\documentstyle[epsf,prb,aps,twocolumn,floats]{revtex}

\begin{document}
\draft
\wideabs{

\title{Metal-insulator transition induced by $^{16}$O --$^{18}$O oxygen isotope
exchange in colossal negative magnetoresistance manganites}
\author{N. A. Babushkina, L. M. Belova\cite{kiae} and V. I. Ozhogin}
\address{Russian Research Center ``Kurchatov Institute'', \\
Kurchatov sqr. 1, Moscow, 123182 Russia}
\author{O. Yu. Gorbenko, A. R. Kaul and A. A. Bosak}
\address{Chemistry Department, Moscow State University, \\
Vorobievy Gory, Moscow, 119899 Russia}
\author{D. I. Khomskii}
\address{Materials Science Center, University of Groningen,\\
Nijenborgh 4, 9747 AG Groningen, The Netherlands\\
Lebedev Physics Institute, Leninskii pr. 53, Moscow, 117924 Russia}
\author{K. I. Kugel}
\address{Scientific Center for Applied Problems in Electrodynamics,\\
Izhorskaya St. 13/19, Moscow, 127412 Russia}
\maketitle

\begin{abstract}
The effect of $^{16}$O~$\rightarrow $~$^{18}$O isotope exchange on the
electric resistivity was studied for (La$_{1-y}$Pr$_{y}$)$_{0.7}$Ca$_{0.3}$%
MnO$_{3}$ ceramic samples. Depending on $y$, this mixed perovskite exhibited
different types of low-temperature behavior ranging from ferromagnetic metal
(FM) to charge ordered (CO) antiferromagnetic insulator. It was found that
at $y=0.75$, the substitution of $^{16}$O by $^{18}$O results in the
reversible transition from a FM to a CO insulator at zero magnetic field.
The applied magnetic field ($H\geq 2$~T) transformed the sample with $^{18}$%
O again to the metallic state and caused the increase in the FM transition
temperature $T_{C}$ of the $^{16}$O sample. As a result, the isotope shift
of $T_{C}$ at $H=2$~T was as high as 63~K. Such unique sensitivity of the
system to oxygen isotope exchange, giving rise even to the metal-insulator
transition, is discussed in terms of the isotope dependence of the effective
electron bandwidth which shifts the balance between the CO and FM phases.
\end{abstract}
}

\section{INTRODUCTION}

The perovskite manganites R$_{1-x}$M$_{x}$MnO$_{3}$, (Ref.~\onlinecite{rao}) 
(R$^{3+}$ is a rare earth cation, M is a doubly charged cation with a large
ionic radius, both R and M occupy A positions of the ABO$_{3}$ perovskite
lattice) attract a considerable current interest owing to the recent
discovery of colossal negative magnetoresistance (CMR). CMR in manganites
stems from the semiconductor-metal transition accompanied by arising of
ferromagnetic ordering, both transitions are usually interpreted on basis of
the double exchange (DE) mechanism.\cite{zener} However, double exchange
alone is insufficient to explain the whole variety of phenomena observed in
manganites. These materials are characterized by a strong interplay of
structural, orbital, and spin degrees of freedom dramatically affecting
their transport properties. The important role of the electron-lattice
interaction was pointed out in Ref.\onlinecite{mil}. There is a number of
factors contributing to these interactions, including the Jahn-Teller nature
of Mn$^{3+}$ ions, the strong dependence of electron transfer on the
Mn--O--Mn bond angle related to the average ionic radius at the A site,\cite
{hw,font} trapping of charge carriers due to the optical breathing mode, and
different polaronic effects.

Another feature of the perovskite manganites is the possible formation of
the charge ordered (CO) state, manifesting itself in localization of charge
carriers accompanied by the regular arrangement of Mn$^{3+}$ and Mn$^{4+}$
ions.\cite{jir,yos,lee} The CO state is usually characterized by
semiconductor-like and antiferromagnetic behavior. The tendency towards CO
is observed with a decrease in the average ionic radius in A position. Such
an ordering, characteristic of Pr$_{1-x}$Ca$_{x}$MnO$_{3}$ ($x=0.2$ -- 0.5),
was actively studied recently.\cite{yos,lee} The charge ordering gives rise
to lattice distortions resulting in the instability of the initial crystal
lattice.

The significant role of electron-lattice interactions in manganites is
confirmed by a pronounced isotope effect on the electrical and magnetic
properties. The effect of $^{16}$O -- $^{18}$O isotope exchange on the
magnetization of La$_{0.8}$Ca$_{0.2}$MnO$_{3}$ was first reported in Refs.%
\onlinecite{zh1,zh2}. The obtained isotope shift of the Curie temperature
was as large as 21~K.

The isotope effect should be even more pronounced in the vicinity of lattice
instability related to the charge ordering. Bearing this in mind, we studied
the effect of oxygen isotope substitution on the electrical resistivity of a
mixed compound (La$_{1-y}$Pr$_{y}$)$_{0.7}$Ca$_{0.3}$MnO$_{3}$. By varying
the relative contents of Pr and La (i.e., changing the average radius of the
rare earth ion) it is possible to obtain the different types of
low-temperature behavior [from the ferromagnetic metal (FM) La$_{0.7}$Ca$%
_{0.3}$MnO$_{3}$ to the charge-ordered antiferromagnetic insulator Pr$_{0.7}$%
Ca$_{0.3}$MnO$_{3}$].\cite{hw,gor}

The isotope effect in electrical resistivity which we observed for La$%
_{0.175}$Pr$_{0.525}$Ca$_{0.3}$MnO$_{3}$ ($y=0.75$) samples appeared to be
far in the excess of our expectations since the $^{16}$O~--~$^{18}$O
exchange resulted not only in the significant lowering of the Curie
temperature, but also in the metal-insulator transition.

\section{EXPERIMENT}

To prepare the ceramic samples, water solutions of La(NO$_{3}$)$_{3}$, Pr(NO$%
_{3}$)$_{3}$, Ca(NO$_{3}$)$_{2}$, and Mn(NO$_{3}$)$_{2}$ were mixed together
in proper ratios. Then ash-free paper filters were soaked with the solution.
Dried at 120${{}^{\circ }}$C, the paper was burned and the residue was
annealed at 700${{}^{\circ }}$C for 2 h. The powder residue was pressed into
the form of pellets that were sintered at 1200${{}^{\circ }}$C for 12 h in
air. X-ray diffraction (XRD) showed that the La$_{0.175}$Pr$_{0.525}$Ca$%
_{0.3}$MnO$_{3}$ ceramic samples produced were of single phase with an
orthorhombic structure (lattice parameters $a=0.5436(2)$~nm, $b=0.5461(2)$%
~nm, $c=0.7686(3)$~nm at 300~K).

The (La$_{1-y}$Pr$_{y}$)$_{0.7}$Ca$_{0.3}$MnO$_{3}$ system appeared to be
convenient for oxygen isotope substitution since its oxygen stoichiometry
depends only slightly on variations in the thermal treatment conditions
(temperature and partial pressure of O$_{2}$).

The method used for oxygen isotope substitution $^{16}$O~--~$^{18}$O in the
selected manganite was similar to that utilized for high-temperature
superconductor ceramics.\cite{bab} Two 1$\times $1$\times $8 mm$^{3}$ bars
were cut from the sintered pellet of La$_{0.175}$Pr$_{0.525}$Ca$_{0.3}$MnO$%
_{3}$ and placed in alumina boats which were then mounted in the furnace
inside two quartz tubes. The quartz tubes were parts of two identical closed
loops with forced circulation of gas. Both samples were treated
simultaneously: one sample was heated in $^{16}$O$_{2}$ atmosphere, the
other sample was heated in $^{18}$O$_{2}$ (the molar fraction of $^{18}$O$%
_{2}$ was 85\%). The diffusion annealing was carried out for 48 h at 950${%
{}^{\circ }}$C under oxygen pressure of 1 bar. The $^{18}$O content in the
samples was determined by measurement of the weight change of the samples
after the isotope enrichment. As a result, the completeness of the isotope
exchange was verified.

The electric resistivity was measured by the standard four-probe technique
at temperatures down to 4.2~K. The highest resistance that could be measured
with the experimental setup was 1~G$\Omega $. The resistivity measurements
in the magnetic field applied along the bar sample were performed only
during the cooling stages. The magnetic properties were not measured in the
current study, thus the assignment of the magnetic states was qualitative
and based on experimental data for similar compounds published earlier.\cite
{hw,jir,yos,lee,kas}

\section{RESULTS}

\begin{figure}[btp]
 \begin{center}
  \leavevmode
  \epsfxsize=0.95\columnwidth \epsfbox {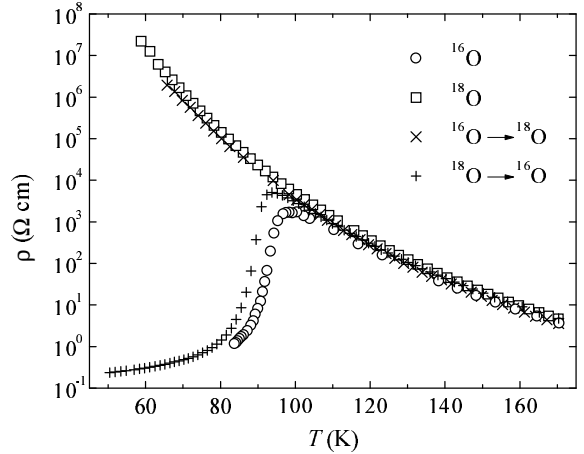}
  \caption{Temperature dependence of resistivity for $^{16}$O and $^{18}$O
samples of La$_{0.175}$Pr$_{0.525}$Ca$_{0.3}$MnO$_3$ before and after oxygen
back-exchange.}
  \label{fig1}
 \end{center}
\end{figure}

The temperature dependence of resistivity $\rho (T)$ in zero magnetic field
for the La$_{0.175}$Pr$_{0.525}$Ca$_{0.3}$MnO$_{3}$ samples is shown in Fig.~%
\ref{fig1}. The resistivity curves for samples treated in $^{16}$O$_{2}$ and 
$^{18}$O$_{2}$ differ drastically. For both the $^{16}$O and $^{18}$O
samples a semiconductor-like activation behavior is observed at high
temperature and the resistivity increases dramatically with decreasing
temperature. Then the $^{16}$O sample exhibited a clearly pronounced
resistivity peak at $T_{C}=95$~K associated with transition to the FM state.
This value of $T_{C}$ is in good agreement with Ref.\onlinecite{hw}. For the 
$^{18}$O sample the resistivity increases monotonically with decreasing
temperature and reaches about 10$^{8}$ $\Omega \cdot $cm at 50~K, below
which the $\rho (T)$ value is so large that it exceeds our measuring limit.
This behavior can be attributed to the charge-ordered state which is usually
responsible for such an increase in the resistivity.\cite{kas}

To make sure that the effect is due precisely to the isotope exchange, we
carried out an isotope back-exchange. The completeness of the exchange was
proved again by change of the sample weight. The sample, which had been
first saturated by $^{18}$O and was insulating thereafter, became metallic
below 95~K after subsequent annealing in $^{16}$O$_{2}$. Correspondingly,
the sample, that had been first treated in $^{16}$O$_{2}$ and was metallic
below 95~K thereafter became an insulator after the subsequent treatment in $%
^{18}$O$_{2}$ (Fig.~\ref{fig1}).

Application of the magnetic field is known to provide melting of the CO
state in (La$_{1-y}$Pr$_{y}$)$_{0.7}$Ca$_{0.3}$MnO$_{3}$ resulting in a
metamagnetic phase transition with a drastic decrease of the electric
resistivity.\cite{hw,gor} This is a magnetic transition of the first order,
exhibiting a pronounced temperature hysteresis. To avoid hysteresis the
samples were heated above 150~K, then cooled in the applied magnetic field.

\begin{figure}[btp]
 \begin{center}
  \leavevmode
  \epsfxsize=0.95\columnwidth \epsfbox {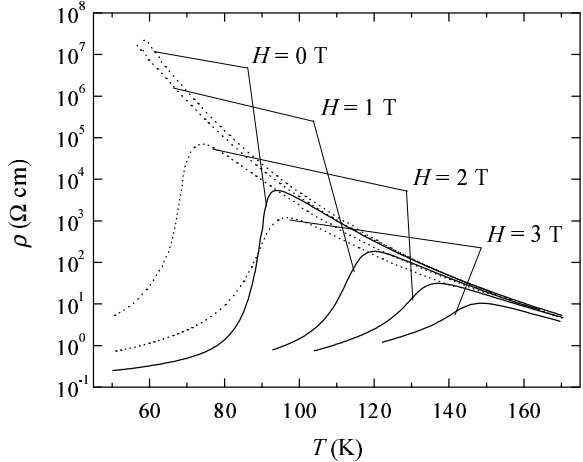}
  \caption{Temperature dependence of resistivity for $^{16}$O and $^{18}$O
samples of La$_{0.175}$Pr$_{0.525}$Ca$_{0.3}$MnO$_3$ in different magnetic
fields. Solid line is for $^{16}$O sample, dotted line is for $^{18}$O
sample.}
  \label{fig2}
 \end{center}
\end{figure}

It is quite interesting to evaluate the stability of the CO state induced by
the isotope exchange to the applied magnetic field. Temperature dependencies
of resistivity $\rho (T)$ in the magnetic field are shown in Fig.~\ref{fig2}.

We found that a magnetic field of 1~T was not enough to suppress the CO
state in La$_{0.175}$Pr$_{0.525}$Ca$_{0.3}$Mn($^{18}$O$_{0.85}$~$^{16}$O$%
_{0.15}$)$_{3}$ in spite of the more than 30~K upward shift of $T_{C}$ in
the case of La$_{0.175}$Pr$_{0.525}$Ca$_{0.3}$Mn($^{16}$O)$_{3}$.
Nevertheless, in a field of 2~T melting of the charge-ordered state was
clearly demonstrated (Fig.~\ref{fig2}).

The isotope shift of $T_{C}$ was as large as 63~K in a field of 2~T and 54~K
in a field of 3~T, which is considerably larger than isotope effects ever
reported for R$_{1-x}$M$_{x}$MnO$_{3}$. It is of note, that $T_{C}$ grows
significantly with the increase of the applied magnetic field for both
samples. The $\rho (T)$ curves at higher temperature are nearly independent
of the isotope exchange (Fig.~\ref{fig2}).

Under an external field of 2~T, the resistivity drops by more than seven
orders of magnitude to ($\rho _{2T}=1.5$ $\Omega \cdot $cm) at 50~K,
exhibiting one of the largest magnitudes of CMR (10$^{7}$). For the $^{18}$O
sample the large decrease in $\rho $, and thus CMR, can be related to the
field-induced melting of the CO state. For the $^{16}$O sample the magnitude
of the CMR is much less (10$^{4}$ for $H=2$~T).

Note the striking difference between the results for CMR in the $^{16}$O and 
$^{18}$O samples. In the $^{16}$O sample metallic conduction and
ferromagnetism occur spontaneously below the Curie temperature $T_{C}$. CMR
arises from the abrupt increase in electron hopping via DE at the onset of
the spontaneous FM ordering. Sizable CMR exists in the vicinity of $T_{C}$.
In the $^{18}$O sample neither metallic conduction nor ferromagnetism is
observed in zero field. CMR originates from the field-induced simultaneous
semiconductor-metal and AF--FM transitions.

\section{DISCUSSION}

Let us discuss qualitatively the possible mechanisms of the observed isotope
effect. Note that $^{16}\text{O}$ -- $^{18}$O exchange stabilizes the
high-resistivity charge-ordered state at zero magnetic field and at the same
time shifts toward lower temperatures the onset of the ferromagnetic metal
phase at sufficiently high applied field. The samples studied in current
research lie close to the FM--CO phase boundary and a relatively small
influence can in principle transform one state into another, this can be
achieved even by isotope substitution.

The relative stability of different phases is determined mainly by the
electron bandwidth, or the effective hopping integral $t_{\text{eff}}$. The CO
state is favored in the situation when $t_{\text{eff}}$ is small enough.\cite
{hw,font} One can show this for a simple case of one electron per two sites,
using the model initially formulated for magnetite,\cite{cul,hom} which
takes into account the Coulomb repulsion $V$ of electrons at neighboring
lattice sites. Its Hamiltonian has the form

\begin{equation}
H=t\sum\limits_{<i,j>}a_{i}^{+}a_{j}+\frac{V}{2}\sum\limits_{<i,j>}n_{i}n_{j}\,,
\label{eq1}
\end{equation}
where $a_{i}^{+}$ and $a_{i}$ are creation and annihilation operators for an
electron at the $i$th site, $n=a^{+}a$, and $<...>$ means the summation over
the nearest neighbors. On-site Coulomb interaction $U$ is assumed to be the
largest parameter, $U\rightarrow \infty $.

The standard mean-field treatment shows \cite{hom} that the CO state occurs
if the hopping integral $t$ is less than a certain critical value, $%
t<t_{c}=Vz/2$ ($z$ is the number of nearest neighbors). The CO critical
temperature is given by the expression

\begin{equation}
T_{CO}\approx Vz\sqrt{1-(t/t_{c})^{2}}  \label{eq2}
\end{equation}
and is reduced with the increase of $t$, especially strongly for $t\approx \
t_{c}$. Thus, the increase of the hopping integral $t$ destabilizes the CO
phase.

The $t_{\text{eff}}$ value can be determined by averaging $t$ over corresponding
lattice vibrations that, in turn, depend on the isotope composition of the
material. In the simplest case: $t(r)\propto \exp (-\alpha r)$ and

\begin{equation}
t_{\text{eff}}=<t>=t_{0}\left( 1+\frac{1}{2}\alpha ^{2}<u^{2}>\right)\,,  
\label{eq3}
\end{equation}
where the interatomic distance $r=r_{0}+u$, $t_{0}=t(r_{0})$, and $\alpha
\propto 1/r_{0}$.

The mean-square displacement $<u^{2}>$, and hence $t_{\text{eff}}$, depend on the
ionic mass $M$ (even at $T=0$ due to the zero-point vibrations).

In case of dominant contribution of zero-point vibrations, we have: 
$$
<u^{2}>=\hbar /\left( 2M\Theta _{D}\right) =\hbar /\left( 2\left( BM\right)
^{1/2}\right)\,,
$$
where $B$ is the bulk modulus. Thus
\begin{equation}
\delta t_{\text{eff}}\approx -t_{0}\left( <u^{2}>/r_{0}^{2}\right) \left( \delta
M/M\right).   
\label{eq4}
\end{equation}
The corresponding change in $t_{\text{eff}}$ is generally not large, especially
taking into account that we have to use the reduced mass instead of the pure
ionic mass. But this change may still be sufficient to shift our system from
one state to another if we are close to the phase boundary at the phase
diagram, which is apparently the case in our system. In this situation $%
t\approx t_{c}$ and, as it follows from Eq.~(\ref{eq2}),
\begin{equation}
\delta T_{CO}\propto \delta t/\sqrt{1-\left( t/t_{c}\right) ^{2}}.
\label{eq5}
\end{equation}
Note that in Eq.~(\ref{eq2}) the typical values of the intersite Coulomb
repulsion $Vz\approx \ 1$~eV (10$^{4}$~K), while $T_{CO}\sim $10$^{2}$~K;
hence the square roots in Eq.~(\ref{eq2}) and Eq.~(\ref{eq5}) are of the 
order of 10$^{-2}$. Thus, $\delta T_{CO}\propto \delta t\cdot 10^{2}$ and 
it is not surprising that even a relatively small variation of 
$t_{\text{eff}}$ caused by isotope substitution leads to such a dramatic 
change in properties.

There are also other factors affecting $t_{\text{eff}}$, such as a possible 
change
of the Mn--O--Mn angle, polaronic effects, and lattice anharmonicity,
especially in the vicinity of the lattice instability that manifests itself
in the system under study. They should be taken into account in a more
detailed analysis.

The experimental results obtained correlate with relationships (\ref{eq4})
and (\ref{eq5}) indicating that $t(M)$ is a decreasing function of $M$ since
the charge-ordered state becomes stable for the heavier isotope. The
behavior of $T_{C}$ at nonzero magnetic fields does not contradict the
character of the $t(M)$ dependency. In fact, if the ferromagnetism stems
from the double exchange (i.e., $t<J_{H}$, where $J_{H}$ is the intraatomic
exchange coupling giving rise to the Hund rule), then $T_{C}$ is
proportional to $t$. For decreasing $t(M)$, we have $T_{C}\left( ^{18}\text{O%
}\right) <T_{C}\left( ^{16}\text{O}\right) $, in agreement with our results
and with earlier experiments on the isotope effect in manganites.\cite
{zh1,zh2} However, it is not quite clear why the isotope effect for $T_{C}$
is so large. In Ref.\onlinecite{kres} this fact is related to manifestations
of the Jahn-Teller effect which can enhance the unharmonic contribution to
polaronic narrowing of $t$. In the vicinity of lattice instability caused by
the possibility of charge ordering this effect can be even more pronounced.
An additional contribution to the isotope shift of $T_{C}$ may be due to the
isotope mass dependence of the fraction of Mn--O--Mn linkages involved in
the double exchange. This fraction is determined by the quantity $\omega
_{R}\exp \left( -\varepsilon _{p}/t\right) $, where $\omega _{R}$ is the
frequency of the optical breathing mode that traps a carrier at a single
cation site and $\varepsilon _{p}$ is the characteristic energy of manganese
clusters participating in the double exchange interactions.\cite{good} We
have $\omega _{R}\sim M_{O}^{-1/2}$ ($M_{O}$ is the oxygen atomic mass),
hence this contribution results in the additional $T_{C}$ lowering at the $%
^{16}$O~$\rightarrow $~$^{18}$O exchange.

\section{CONCLUSIONS}

The electric resistivity of La$_{0.175}$Pr$_{0.525}$Ca$_{0.3}$MnO$_{3}$
ceramics demonstrates very high sensitivity to the oxygen isotope exchange.
The samples with $^{16}$O are metallic at low temperatures, while
substitution of $^{16}$O by $^{18}$O results in the insulator-like behavior.
This transition is completely reversible and the sample returns to the
original metallic state after backsubstitution $^{18}$O~$\rightarrow $~$^{16}
$O. The magnetic field exceeding 1~T restores the metallicity, but the
isotope shift of the resistivity peak is very large, 63 and 54~K at $H=2$%
~and 3~T, respectively. Moreover, the CMR is much more pronounced in the
case of $^{18}$O (the resistivity drops for $^{18}$O and $^{16}$O differ by
a factor attaining the value of 10$^{3}$ at $H=2$~T).

We argue that this enormous isotope effect is caused by modification of the
effective hopping integrals and the resulting electron bandwidth due to
isotope substitution, which shifts the relative stability of the CO versus
FM states and leads to a transition between these phases. Simple model
considerations suggest a significant enhancement of the isotope effect near
the onset of the charge-ordered state and the corresponding lattice
instability. To analyze this problem quantitatively we must take into
account a whole set of competing mechanisms: electron-phonon interaction,
polaronic band narrowing, the Jahn-Teller effect, double exchange, intersite
Coulomb repulsion of electrons, etc. Nonetheless, we think that our simple
arguments correctly describe the main physics of the phenomenon observed.

\acknowledgements 

The authors are grateful to A.~N.~Taldenkov and A.~V.~Inyushkin for help in
the work. The work was partially supported by the Russian Foundation for
Basic Research, Project No.~97-03-32979a. One of the authors (D.~I.~K.)
acknowledges The Netherlands Foundation for the Fundamental Research of
Matter (FOM) for support.

\end{document}